\documentclass[preprint, 2023]{{PHMSociety}}
\usepackage{graphicx}
\usepackage{amsmath}

\usepackage{algorithm}
\usepackage{algpseudocode}
\usepackage{gensymb}
\usepackage{multirow}
\usepackage{bm}
\usepackage{amssymb}
\begin{document}
\title{A Comparison of Residual-based Methods on Fault Detection}

\author{%
	Chi-Ching Hsu\authorNumber{1}, Gaetan Frusque\authorNumber{2}, and Olga Fink\authorNumber{3}}

\address{
	\affiliation{{1}}{High Voltage Laboratory, ETH Zurich, Zurich, Switzerland}{ 
		{\email{hsu@eeh.ee.ethz.ch}} 
		} 
	\tabularnewline 
	\affiliation{{2,3}}{Intelligent Maintenance and Operations Systems Laboratory, EPFL, Lausanne, Switzerland}{ 
		{\email{gaetan.frusque@epfl.ch}}\\
        {\email{olga.fink@epfl.ch}}
		} 
}

\maketitle


\begin{abstract}
An important initial step in fault detection for complex industrial systems is gaining an understanding of their health condition. Subsequently, continuous monitoring of this health condition becomes crucial to observe its evolution, track changes over time, and isolate faults. As faults are typically rare occurrences, it is essential to perform this monitoring in an unsupervised manner. Various approaches have been proposed not only to detect faults in an unsupervised manner but also to distinguish between different potential fault types. In this study, we perform a comprehensive comparison between two residual-based approaches: autoencoders, and the input-output models that establish a mapping between operating conditions and sensor readings. We explore the sensor-wise residuals and aggregated residuals for the entire system in both methods. The performance evaluation focuses on three tasks: health indicator construction, fault detection, and health indicator interpretation.
To perform the comparison, we utilize the Commercial Modular Aero-Propulsion System Simulation (C-MAPSS) dynamical model,  specifically a subset of the turbofan engine dataset containing three different fault types. All models are trained exclusively on healthy data. Fault detection is achieved by applying a threshold that is determined based on the healthy condition. The detection results reveal that both models are capable of detecting faults with an average delay of around 20 cycles and maintain a low false positive rate. While the fault detection performance is similar for both models, the input-output model provides better interpretability regarding potential fault types and the possible faulty components. 

\end{abstract}

\section{Introduction}
\label{sec:intro}
Determining the health state of complex industrial systems, such as turbofan engines, under different operating conditions has become feasible due to the abundance of condition monitoring data collected by diverse sensors. A health state is usually described by a health indicator or a condition indicator, which is a value that reflects system health conditions and health status in a predictable way as a system degrades~\cite{lei2018machinery}. In complex systems, inferring these indicators and monitoring their evolution over time provide a more comprehensive understanding of the system's health and enable effective condition monitoring.


Typically, a distinction is made between condition indicators and health indicators. A condition indicator refers to a specific feature within system data that exhibits predictable changes as the system undergoes degradation or operates in different operational modes~\cite{fink2020potential}. It encompasses any feature that proves valuable in differentiating normal operation from faulty or any deviation from normal operation. Health indicators, in contrast, integrate multiple condition indicators into a single value, providing the end user with a comprehensive health status of the component.

Different approaches have been proposed to extract and learn the condition and health indicators of a system. These approaches can be categorised into three main categories: feature-based, one-class classification-based (OCC-based), and residual-based methods. Feature-based methods primarily focus on condition indicators. These methods identify relevant features that exhibit predictable changes as the system deteriorates, and they detect early-stage faults by directly applying the threshold method to the feature values. For instance, the relative root mean square (RMS) value of the acceleration signals from bearings serves as an indicator of wear evolution~\cite{pan2020two}. While this approach is straightforward, it requires expert knowledge and can be sensitive to varying operating conditions~\cite{saufi2019challenges}. 

While the feature-based methods for extracting condition indicators focus on expert-based determination of one or several features that capture the condition evolution of different components, OCC-based methods focus on learning a global indicator that represents the health state of the systems~\cite{michau2020feature}. OCC-based methods are particularly suitable for setups of missing faulty samples during training. They are trained on data from one class (usually healthy data). While OCC outputs can provide binary health information (healthy or unhealthy), measuring the distance to the healthy data can effectively infer the evolution of the degradation. This distance can be interpreted as a health indicator, which can also be derived for subsystems by considering a subset of condition monitoring signals related to the specific subsystem. It can then be utilized to monitor the evolution of health conditions, detect anomalies, or distinguish between different severity levels of faults ~\cite{michau2017deep}. 

The third direction encompasses residual-based methods, which extract health indicators based on the residuals. The residuals are the differences between the measured values and the predicted outputs, serving as indicators of any deviation from the healthy training dataset~\cite{chao2019hybrid}. These methods can be categorised into two main types: autoencoders and input-output models. Autoencoders are trained to reconstruct their own inputs, whereas input-output models establish mappings between operating conditions and sensor readings. For example, in the case of a turbofan engine, operating conditions are used as inputs to the full authority digital electronic control (FADEC) to derive monitored signals as health indicators~\cite{rausch2007integrated}. Both, input-output models and autoencoders are typically trained solely on healthy data and, as a result, learn the healthy data distribution. Consequently, when presented with anomalous samples stemming from a different data distribution, they generate significant residuals.

In residual-based methods, there are various approaches to calculating residuals, particularly for aggregating the residuals of multivariate condition monitoring signals. One of the most commonly used methods for aggregating residuals is to compute their sum, offering a comprehensive representation of the overall global health condition~\cite{guo2022unsupervised}. Another approach to utilizing the residuals is to bypass their aggregation and instead use them individually. By analyzing the residuals individually, it becomes possible to identify the specific signals most affected by each fault type~\cite{reddy2016anomaly, michau2020feature}. This approach enables fault segmentation and fault diagnostics, as different faults tend to impact distinct sets of signals.

However, since residual-based models are trained solely on healthy data and residuals are calculated based on the distance to the training data distribution, they are unable to differentiate between a fault and a new operating condition. 
In other words, high residuals may not necessarily indicate deteriorating health conditions of a system but rather the presence of a novel operating condition. This presents a significant challenge in accurately inferring the health state or conducting further downstream tasks, such as fault detection and fault segmentation. 




While several residual-based approaches have been applied to different case studies~\cite{chao2019hybrid, lovberg2021remaining, darrah2022developing}, to the best of our knowledge, their performances have not been compared. In this study, we compare two residual-based methods: autoencoders and input-output models. We use a simulated turbofan dataset with three different fault engine components exhibiting degradation behavior. We evaluate their performance by first constructing the health condition using two types of residuals as health indicators. Subsequently, we perform fault detection and interpret the constructed health indicators.

\section{Method}
\label{sec:methods}
\begin{figure*}[htbp]
\centerline{\includegraphics[width=\textwidth]{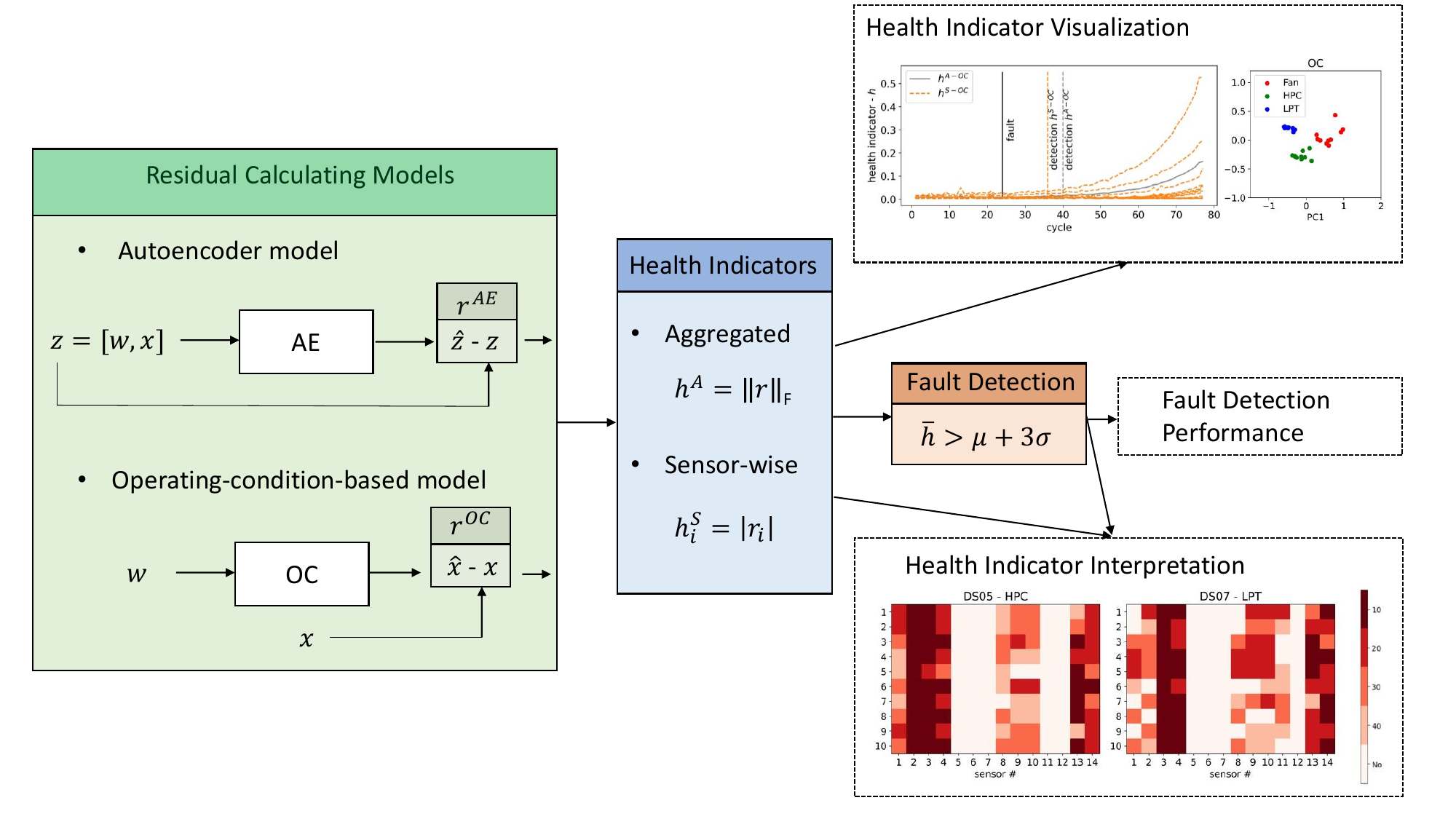}}
\caption{Overall architecture of the testing framework. The framework includes residual calculating models, health indicator construction, and fault detection algorithm. We assess the performances of each health indicators based on detection performances, data visualization and the interpretive capability associated with the machine's condition.}
\label{fig:flowchart}
\end{figure*}

We present in this section the overall proposed testing framework as summarised in Figure~\ref{fig:flowchart}. First, in Section~\ref{sec:hi_models} we present two strategies for calculating residuals, enabling us to identify instances when the data distribution deviates from the healthy distribution. Second, we describe how the residuals can be used to construct health indicators in Section~\ref{sec:hi}. We show in Section~\ref{sec:fault_detection} how we infer the fault initiation from the constructed health indicators.

\subsection{Residual Calculating Models}
\label{sec:hi_models}
\subsubsection*{Autoencoder Model (AE Model)}
\label{sec:hi_ae}
One commonly used residual-based models is the autoencoder. It aims to encode inputs into latent space with the encoder $E_{\theta_e}(\bullet)$ while preserving important information, and then decode it back to its original form using the decoder $D_{\theta_d}(\bullet)$. Here, $\theta_e$ and $\theta_d$ represent the model parameters of the encoder and decoder, respectively.
In our case, we consider a multivariate dataset containing several sensors $\bm{z}_t \in \mathbb{R}^{N_z}$, where $t$ is the time index and $N_z$ is the number of sensors. To learn the distribution of the healthy samples, the autoencoder is trained exclusively on samples captured during the early stages of the system's lifecycle, denoted as $t \in \{1,...,T_H\}$. In this period, we assume that the system is in a healthy state. 
We denote by $\bm{r}^{ae}$ the residual of the AE model, which represents the difference between the output and input signal. Mathematically, it can be written as: 
\begin{equation}
\label{eq:res_ae}
    \bm{r}^{\rm AE}_t =  \bm{z}_t - D_{\theta_d}(E_{\theta_e}(\bm{z}_t)).
\end{equation}
By training the autoencoder, we aim to find the parameters $\theta_e$ and $\theta_d$ that minimise the residuals in terms of mean square error. With $|| \bullet ||_F$ the Frobenius norm, the optimisation problem of the autoencoder model can be written as: 
\begin{align}\label{eq:loss_ae}
&\underset{\theta_e, \theta_d }{\rm argmin} \hspace{0.4cm}  \frac{1}{T_H} \sum_{t=1}^{T_H} \mid\mid  \bm{r}^{AE}_t \mid\mid_F.   
\end{align}

\subsubsection*{Operating-conditions-based Model (OC Model)}
In addition to the autoencoders, we also evaluate an input-output method that maps the operating conditions to the sensor readings~\cite{lovberg2021remaining, darrah2022developing}. We refer to this model as the operating-conditions-based model (OC Model). The OC model is based on operating condition descriptors that characterize the state of a system. For instance, in an industrial bearing, these descriptors include rotating speed and static loading. For a turbofan engine, the considered state descriptors include altitude, flight Mach number, throttle-resolver angle, and the total temperature at the engine fan inlet. 
The multivariate time series $\bm{z}_t$ can be subdivided into operating condition descriptors $\bm{w}_t \in \mathbb{R}^{N_w}$ and sensor readings $\bm{x}_t \in \mathbb{R}^{N_x}$. Here, $N_w$ and $N_x$ represent the number of operating condition descriptors and sensor readings, respectively, and $N_z=N_w+N_x$.
The OC model, denoted as $M_{\theta_m}(\bullet)$, aims to establish a mapping between the operating conditions and sensor readings by learning the functional relationship between the two. The OC Model is expected to be more robust to variations in operating conditions compared to the autoencoders where the operating condition descriptors are part of the reconstructed signals. 
We define the residual of the OC Model as the difference between the estimated and real sensor readings such as: 
\begin{equation}
\label{eq:res_oc}
    \bm{r}^{\rm OC}_t =  \bm{x}_t - M_{\theta_m}(\bm{w}_t),
\end{equation}

The OC model is trained by finding the parameters $\theta_m$ that minimise the residuals. The corresponding optimisation problem can be written as follows: 

\begin{align}\label{eq:loss_oc}
&\underset{\theta_m }{\rm argmin} \hspace{0.4cm}  \frac{1}{T_H} \sum_{t=1}^{T_H} \mid\mid  \bm{r}^{\rm OC}_t \mid\mid_F.   \\
\end{align}

\subsection{Health Indicators}
\label{sec:hi}
In this work, we consider the residuals defined in Equation~\ref{eq:res_ae} and Equation~\ref{eq:res_oc} as a basis for computing the health indicators. We assume that the training dataset is representative of all the operating conditions. Consequently, changes in operating conditions will not be detected as anomalies and an increase in the magnitude of the residuals will be associated  with faulty system conditions. 

We first consider two aggregated health indicators, denoted as $\bm{h}^{\rm A{\text -}AE}$ and $\bm{h}^{\rm A{\text -}OC}$, which represent the norm of the residuals for the AE and OC models, respectively. These indicators combine the residual information from each sensor and can be written at any time $t$ as follows: 
\begin{equation}
\label{eq:residual_ae}
    h^{\rm A{\text -}AE}_t = \mid\mid  \bm{r}^{\rm AE}_t \mid\mid_F,
\end{equation}
\begin{equation}
\label{eq:residual_oc}
    h^{\rm A{\text -}OC}_t = \mid\mid  \bm{r}^{\rm OC}_t \mid\mid_F.
\end{equation}

We also propose two sensor-wise multivariate health indicators, denoted as $\bm{h}^{\rm S{\text -}AE}$ and $\bm{h}^{\rm S{\text -}OC}$. These indicators correspond to the absolute residuals of the AE and OC models, respectively. By considering sensor-wise information, we aim to have indicators that are easier to interpret and more precise for the fault detection task. Using the absolute value operator $|\bullet |$, the health indicators  $\bm{h}^{\rm S{\text -}AE}$ and $\bm{h}^{\rm S{\text -}OC}$ can be written at any time $t$ and for sensor $i$ as follows: 
\begin{equation}
\label{eq:residual_ae_sr}
    h^{\rm S{\text -}AE}_{it} = \mid  r^{\rm AE}_{it} \mid,
\end{equation}
\begin{equation}
\label{eq:residual_oc_sr}
    h^{\rm S{\text -}OC}_{it} = \mid  r^{\rm OC}_{it} \mid.
\end{equation}

\subsection{Fault Detection}
\label{sec:fault_detection}
We propose to use a fault detection algorithm based on a threshold determined by the reconstruction performance of the models on the healthy validation dataset. Considering any of the previously presented health indicators $\bm{h}$, we define the mean $\mu_i$ and standard deviation $\sigma_i$ characterising the healthy condition for sensor $i$ as follows:
\begin{equation}
    \mu_i = \frac{1}{T_H}\sum_{t=1}^{T_H}{h_{it}},
\end{equation}
\begin{equation}
    \sigma^2_i = \frac{1}{T_H}\sum_{t=1}^{T_H}{(h_{it}-\mu_i)^2}.
\end{equation}
Note that for an aggregated health indicator, there is only one set of statistics $\mu$ and $\sigma$, that needs to be computed. 
Thus, we can define the threshold $\tau_i$ for sensor $i$ as:
\begin{equation}
    \tau_i = \mu_i + 3\sigma_i.
\end{equation}

We also divide the time index into $C$ cycles. A cycle is denoted as $n_c$, and it corresponds to a series of time indices $t \in \{T_0, T_{0}+1,...,T_{c-1}-1 \}$ that is a segment of the time samples. A cycle can correspond to a full rotation of a bearing or the flight duration of a turbofan engine. 
The average health indicator during cycle $n_c$ for sensor $i$ is denoted as $\bar{h}_i(n_c)$ and is calculated as follows: 
\begin{equation}
    \bar{h}_i(n_c) = \frac{1}{T_{c+1}-T_{c}}\sum_{t=T_c}^{T_{c+1}}{h_{it}}
\end{equation}
To avoid false alarms, we introduce the waiting cycle number $N_{\rm wait}$. The fault is detected and the alarm is raised only when, for at least one sensor $i$, the corresponding averaged health indicator $\bar{h}_i(n_c)$ is larger than the threshold $\tau_i$ for $N_{\rm wait}$ consecutive cycles. For convenience, we denote $n_0$ as the cycle where the fault is detected and the alarm is raised.

\section{Case Study}
\label{sec:experiment}
\begin{figure}[htbp]
\centerline{\includegraphics[width=80mm]{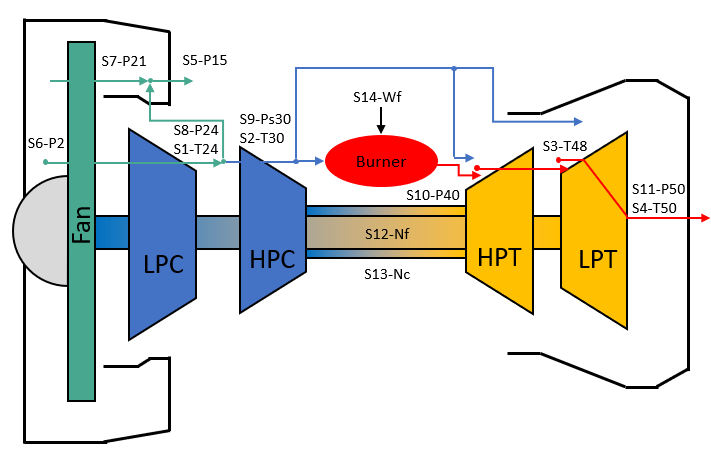}}
\caption{C-MAPSS model schematic representation with sensor position within the engine, adapted from~\protect\cite{arias2021aircraft}}
\label{fig:cmapss}
\end{figure}

\begin{figure}[htbp]
\centerline{\includegraphics[width=80mm]{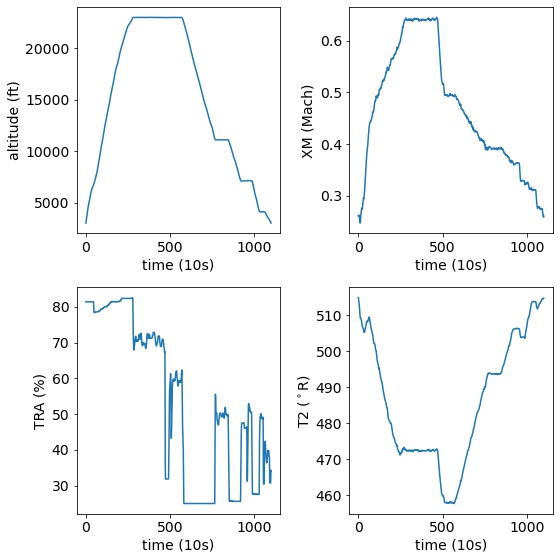}}
\caption{Example of operating condition descriptors $\bm{w}$ for the first cycle of unit 1 in DS04 including altitude , flight Mach number, throttle-resolver angle, and total temperature at the fan inlet, downsampled by a factor of 10}
\label{fig:example_unit1_cycle1_w}
\end{figure}

The dataset we used to evaluate the effectiveness of our proposed approach is the N-CMAPSS~\cite{arias2021aircraft}. This dataset was synthetically generated from the Commercial Modular Aero-Propulsion System Simulation (C-MAPSS) dynamical model. The N-CMAPSS dataset incorporates real flight operating conditions recorded from commercial jets as inputs to the simulation model. It consists of 14 sensor readings $\bm{x}$, which are shown in Figure~\ref{fig:cmapss} with six main components: fan, low pressure compressor (LPC), high pressure compressor (HPC), low pressure turbine (LPT), high pressure turbine (HPT), and burner~\cite{frederick2007user, arias2021aircraft}. Faults are artificially introduced in simulation during the flights. In addition to the sensor readings, the dataset also provides four operating condition descriptors $\bm{w}$ which describes the state of the flight. Figure~\ref{fig:example_unit1_cycle1_w} presents  an instance of a unit for one flight cycle, illustrating these four descriptors. While  certain  research studies incorporate  two additional auxiliary descriptors, namely flight class and positional variable ~\cite{lovberg2021remaining, darrah2022developing},  we have chosen to exclusively utilize the original four operating condition descriptors. 
The descriptions of the sensor reading $\bm{x}$ and the operating condition descriptors $\bm{w}$ are presented in Table~\ref{tb:sensor}. 



The entire dataset is partitioned into multiple sub-datasets, each comprising run-to-failure trajectories of several units affected by distinct fault types. In this work, we focus on sub-datasets DS04, DS05, and DS07. These sub-datasets are chosen because their units are impacted by fault types that affect only a single component, rendering them well-suited for evaluating fault segmentation performance. Other subsets contain units affected by fault types that involve multiple components. Specifically, the fault component for DS04 is the fan, for DS05, it is the HPC and for DS07, it is the LPT. Each sub-dataset contains 10 turbofan engines with the same fault types.



\begin{table}[h]
\begin{center}
\begin{tabular}{ |c|c|c|c| } 
 \hline
\# & Symbol & Description & Units \\ 
 \hline
 \multicolumn{4}{|c|}{sensor readings $\bm{x}$} \\
 \hline
 1 & T24 & Total temperature at LPC outlet & \degree R \\ 
 2 & T30 & Total temperature at HPC outlet & \degree R \\ 
 3 & T48 & Total temperature at HPT outlet & \degree R \\ 
 4 & T50 & Total temperature at LPT outlet & \degree R \\ 
 5 & P15 & Total pressure in bypass-duct & psia \\ 
 6 & P2 & Total pressure at fan inlet & psia \\ 
 7 & P21 & Total pressure at fan outlet & psia \\ 
 8 & P24 & Total pressure at LPC outlet & psia \\ 
 9 & Ps30 & Static pressure at HPC outlet & psia \\ 
 10 & P40 & Total pressure at burner outlet & psia \\ 
 11 & P50 & Total pressure at LPT outlet & psia \\ 
 12 & Nf & Physical fan speed & rpm \\ 
 13 & Nc & Physical core speed & rpm \\ 
 14 & Wf & Fuel flow & pps \\ 
 \hline
 \multicolumn{4}{|c|}{operating condition descriptors $\bm{w}$} \\
 \hline
 15 & alt & altitude & ft \\
 16 & XM & Mach number & - \\
 17 & TRA & throttle-resolver angle & \% \\
 18 & T2 & total temperature at the fan inlet & \degree R \\ 
 \hline
\end{tabular}
\end{center}
\caption{\label{tb:sensor}Sensor readings $\bm{x}$ and operating condition descriptors $\bm{w}$ in the N-CMAPSS dataset with its description and corresponding units}
\end{table}


\subsection{Pre-processing}

All sensors undergo a downsampling process by a factor of 10 to reduce data size and computational costs.
Each sensor reading and each descriptor are standardised to have a zero mean and unit standard deviation. This standardisation process is carried out on the training set and the resulting parameters are then applied to the test and validation sets. The training, validation, and test setup is explained in Section~\ref{sec:trainig_setup}. For this study, we solely consider the cruising phase of the flight as it exhibits a more stable behavior in comparison to the take-off or landing phases. The cruising phase is defined when the normalised flight altitude exceeds 0.85. Normalised flight altitude is calculated by dividing all altitude values by the highest altitude within this cycle. The fault detection waiting cycle $N_{wait}$ is fixed at 3 cycles. 

\subsection{Applied Neural Network Architectures}
For the OC model, we use two 128-neuron layers. For the AE model, we consider three hidden layers with 128-8-128 neurons each. All activation functions used in the models are rectified linear units (ReLU), except for the final layer of both models, which employ a linear activation function.

\subsection{Training Setup}
\label{sec:trainig_setup}
The training was performed for 70 epochs with a batch size of 64, an early stopping waiting epoch of 10, and using Adam optimizer~\cite{kingma2014adam} with $\beta_1 = 0.9$ and $\beta_2 = 0.999$, learning rate of 0.001. We arbitrarily selected the first 16 cycles of each unit to be the healthy data for training. 
The models are trained using healthy data from all 30 units from DS04, DS05, and DS07. The remaining cycles are then assigned to the test set for evaluation. Within the training set, we randomly select 15\% as a validation set for deciding an early-stopping training epoch. 
For each setting, we train the models 5 times with the validation set randomly split. And the results are presented as the average over the 5 realisations.

\subsection{Evaluation Metrics}

To assess the fault detection results of a single engine or unit $u$, we consider the detection delay $d_u$ that can be computed as the difference between the ground truth occurring fault cycle $n_{\rm true}$ and the cycle that raises the alarm $n_0$. It is written as:


\begin{equation}
\label{eq:de_single}
    d_u = n_{\rm true}-n_0
\end{equation}
In case the detection delay is negative ($d_u<0$), it corresponds to a false positive alarm. 
An effective detection algorithm should avoid generating false positive alarms, as they lead to the unnecessary consumption of resources. As a second metric to evaluate the fault detection algorithm, we propose the false positive rate (FPR). FPR corresponds to the number of units with negative detection relative to all units. 

Additionally, the silhouette score~\cite{rousseeuw1987silhouettes} is applied to evaluate the clustering results for fault segmentation. The score measures the similarity of a sample to its own cluster and other clusters. It is calculated for a sample $\rm k$ using  the mean intra-cluster distance $d_{\rm{intra}, k}$ and the mean nearest-cluster distance $d_{\rm nearest, k}$ and defined as follows:

\begin{equation}
    s_k = \frac{d_{\rm nearest, k}-d_{\rm intra, k}}{\rm{max}(d_{\rm intra, k}, d_{\rm nearest, k})}.
\end{equation}
The silhouette score is 1 if the clusters are well separated, 0 if they are overlapped, and -1 if at least one cluster is similar to the others. We take the mean of the scores over all samples.

\section{Results}
\subsection{Health Indicators}

\label{sec:results}
\begin{figure}[htbp]
\centerline{\includegraphics[width=80mm]{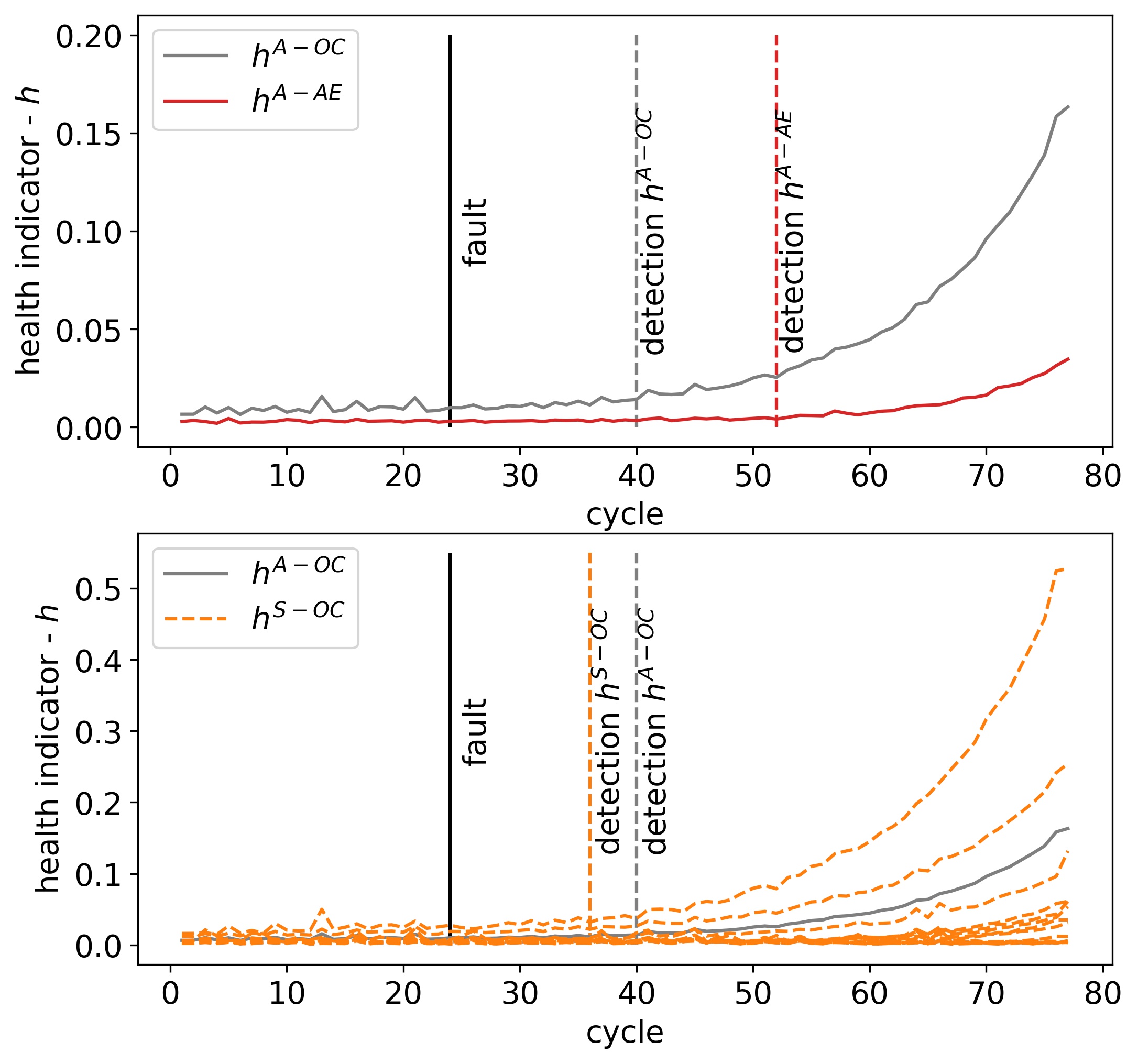}}
\caption{Health indicators calculated from aggregated residuals $\bm{h}^{\rm A}$ obtained from the OC and AE models: $\bm{h}^{\rm A\text{-}OC}$ and $\bm{h}^{\rm A\text{-}AE}$ for DS07 unit 7 (top). Health indicators, aggregated and sensor-wise residuals: $\bm{h}^{\rm A\text{-}OC}$ and $\bm{h}^{\rm S\text{-}OC}$ obtained from the OC model for the same unit (bottom). The vertical black line indicates the fault initiation at cycle 24.}
\label{fig:health_indicators}
\end{figure}

The aggregated health indicators $\bm{h}^{\rm A}$ obtained from both models $\bm{h}^{\rm A\text{-}OC}$ and $\bm{h}^{\rm A\text{-}AE}$ for DS07 unit 7 are shown at the top of Figure~\ref{fig:health_indicators}. This unit is randomly selected for visualization purposes. The fault occurs at cycle 24 and is detected when the fault detection algorithm is applied to $\bm{h}^{\rm A\text{-}OC}$ at cycle 40 and $\bm{h}^{\rm A\text{-}AE}$ at cycle 52. Both health indicators remain constant for approximately 10 to 15 cycles even after the fault occurs. This could be because the fault initially starts with mild severity, but as time progresses, it deteriorates and the health indicator increases at a faster rate. The bottom of Figure~\ref{fig:health_indicators} displays both OC model health indicators $\bm{h}^{\rm A\text{-}OC}$ and $\bm{h}^{\rm S\text{-}OC}$. The sensor-wise residuals $\bm{h}^{\rm S\text{-}OC}$ exhibit different degradation rates for different sensors. Furthermore, some trajectories show exponential behavior, increasing faster than the aggregated $\bm{h}^{\rm A\text{-}OC}$ health indicator. This indicates that specific sensors exhibit the fault behavior before others.

\subsection{Fault Detection Performance}

We evaluate the fault detection performance using the proposed fault detection algorithm on the proposed health indicators. The detection delay $d_u$ of each unit, the average detection delay of each model, and FPR are provided in Table~\ref{tb:fault_detection}. 
On average, the aggregated health indicators $\bm{h}^{\rm A}$ raise an alarm 24.2 cycles and 33.4 cycles after fault initiation for the OC and AE models, respectively. In this case, the FPR is null, indicating that these indicators are robust against false alarms. However, no faults are detected for unit 3 of the DS04 dataset. 
The sensor-wise health indicators $\bm{h}^{\rm S}$ raise alarms at earlier cycles, with an average detection at 15.5 cycles for the OC model and  17.3 cycles for the AE model. The detection occurs earlier than in the aggregated health indicators, primarily due to specific sensors that exhibit faulty behavior first. However, the sensor-wise health indicators are more sensitive to false alarms, as the FPR is not null.

\begin{table}[ht]
\begin{center}
\begin{tabular}{|c|ccc|ccc|}
\hline
     & \multicolumn{3}{c|}{OC}                                  & \multicolumn{3}{c|}{AE}                                  \\ \hline
Unit \#& \multicolumn{1}{c}{DS04} & \multicolumn{1}{c}{DS05} & DS07   & \multicolumn{1}{c}{DS04} & \multicolumn{1}{c}{DS05} & DS07   \\ \hline
\multicolumn{7}{|c|}{$\bm{h}^{\rm S}$} \\\hline
1    & 14.6                     & 9.4                    & 11.8 & 27.4                    & 4.8                     & 9.8  \\
2    & 19.2                    & 14.4                    & 23.8 & 20.6                    & 16.8                    & 23.2 \\
3    & 42.0                    & 7.6                    & 16.8 & 39.4                    & 11.6                     & 17.2 \\
4    & 19.4                    & 3.6                    & 11.2 & 24.8                    & 7.8                    & 7.6  \\
5    & 27.2                    & 6.6                    & 16.2 & 39.0                    & 12.0                     & 11.8  \\
6    & 21.8                    & 8.4                     & 15.4 & 23.2                    & 13.6                     & 13.2 \\
7    & 30.4                    & 12.6                    & 17.2 & 33.0                    & 21.2                    & 19.6 \\
8    & 24.2                    & 3.6                    & 13.0 & 31.0                    & 0.6                     & 11.2  \\
9    & 18.0                    & 12.2                    & 23.0 & 20.6                    & 15.2                    & 18.4 \\
10   & 20.2                    & -0.4                    & 1.0  & 27.2                    & -4.2                    & 0.0 \\ \hline
avg & \multicolumn{3}{|c|}{15.5} & \multicolumn{3}{c|}{17.3} \\\hline
FPR & \multicolumn{3}{|c|}{3.3\%} & \multicolumn{3}{c|}{3.3\%} \\\hline
\multicolumn{7}{|c|}{$\bm{h}^{\rm A}$} \\\hline
1    & 16.2                     & 19.6                    & 17.2   & 48.2                    & 17.8                    & 20.0 \\
2    & 26.8                    & 24.6                   & 32.4  & 34.0                    & 33.0                   & 41.6 \\
3    & -                    & 12.0                  & 25.2  & 67.2                    &   20.6                  & 30.4 \\
4    & 27.0                    & 18.6                    & 16.4  & 33.2                    & 28.6                    & 20.2 \\
5    & 48.4                   & 21.8                    & 21.4  & 59.6                  & 31.8                    & 25.6 \\
6    & 38.6                    & 11.2                     & 21.8  & 39.4                    & 26.4                     & 30.0 \\
7    & 43.2                    & 22.4                    & 20.6  & 45.4                    & 34.0                     & 32.8 \\
8    & 42.6                    & 14.0                    & 15.4  & 48.6                     & 21.0                    & 24.6 \\
9    & 35.4                    & 20.8                    & 30.6   & 31.6                     & 28.6                     & 45.4 \\
10   & 35.4                    & 7.6                   & 13.2 & 43.8                    & 16.0                   & 23.6 \\ \hline
avg & \multicolumn{3}{|c|}{24.2} & \multicolumn{3}{c|}{33.4} \\\hline
FPR & \multicolumn{3}{|c|}{0\%*} & \multicolumn{3}{c|}{0\%} \\\hline
\end{tabular}
\end{center}
\caption{\label{tb:fault_detection}Overview of the fault detection delays $d_u$ using OC and AE models, average over five realisations. "-" means no fault is detected. In total, there are 30 different units, 10 units from each sub-dataset. *: FPR is 0\% but the fault is not detected in one unit}
\end{table}

\subsection{Sensor-wise Health Indicator Visualization}

\begin{figure*}[htbp]
\centerline{\includegraphics[width=0.85\textwidth]{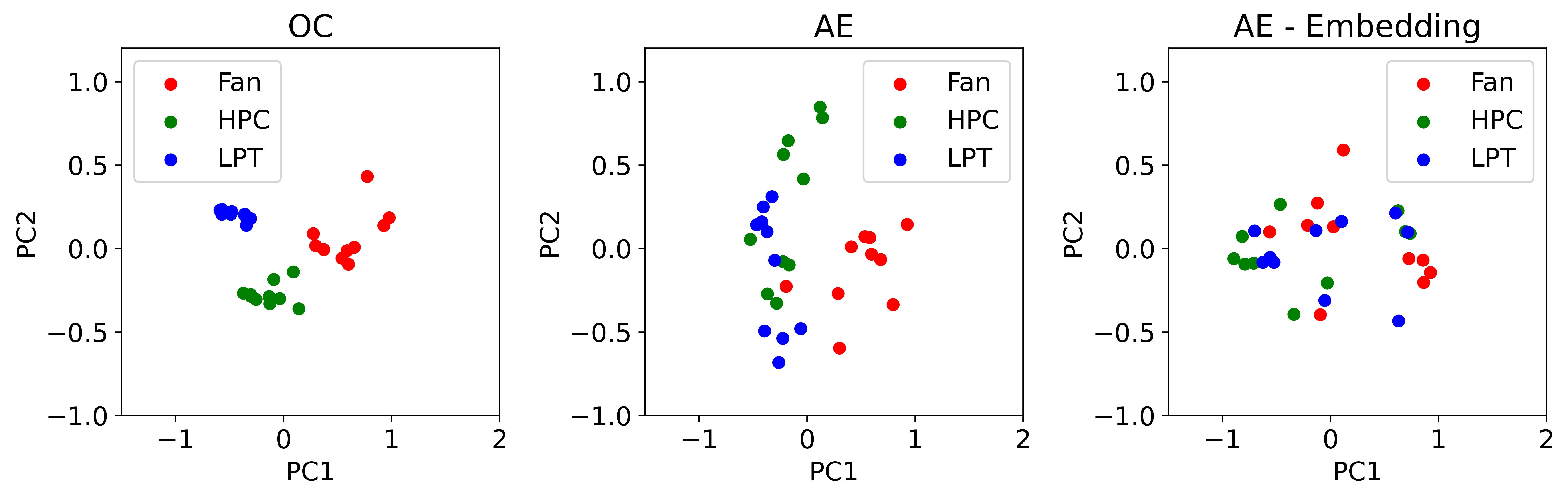}}
\caption{Visualization of the clustering results using the normalised sensor-wise health indicator with $n_c$ set to 10 cycles after fault detection,  color-coded by fault component.}
\label{fig:cluster_visualization_fault}
\end{figure*}

We visualize the normalised sensor-wise health indicator of each unit in low-dimensional space using the first two principal components (PC1 and PC2) from Principal Component Analysis (PCA) in Figure~\ref{fig:cluster_visualization_fault}. The visualization is performed at 10 cycles after the fault is detected ($n_0 + 10$). The value of 10 cycles is chosen to strike a balance, avoiding reaching the end-of-life while ensuring that the fault behavior is exhibited by multiple sensors, rather than just one.
The colors in the visualization are assigned based on the ground-truth fault type, which is not available in reality. In the left figure representing the OC model, units with different fault types form distinct clusters. However, in the case of the AE model, the faults are mixed, and the clusters do not align with specific fault types. As an alternative to the sensor-wise health indicator, we also propose considering the visualization of the output from the embedding latent space layer of the AE model, referred to as AE-Embedding. However, in this case, units with similar fault types do not form coherent clusters. 

\begin{figure}[htbp]
\centerline{\includegraphics[width=55mm]{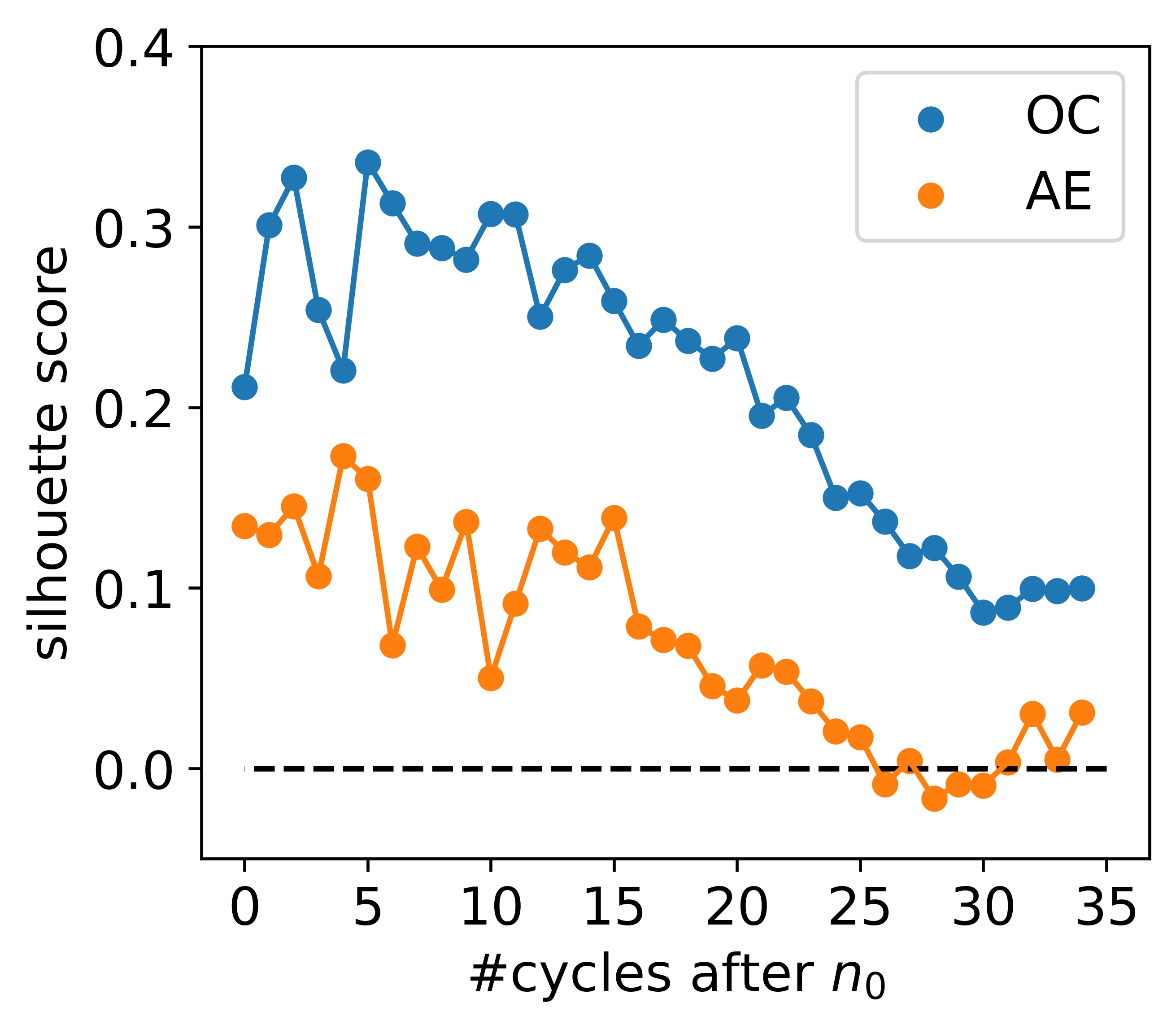}}
\caption{Silhouette scores of $\bm{h}^{\rm S-{OC}}$ and $\bm{h}^{\rm S-{AE}}$ calculated from 0 to 34 cycles after fault detection. The higher silhouette scores suggest more distinct clusters.}
\label{fig:silhouette_scores}
\end{figure}

The silhouette scores with different numbers of cycles after the fault is detected are plotted in Figure~\ref{fig:silhouette_scores} when using both sensor-wise health indicators $\bm{h}^{\rm S-{OC}}$ and $\bm{h}^{\rm S-{AE}}$ as inputs. Using the OC model, the silhouette score is consistently higher than when using the AE model for all cycles, indicating better clustering results. This superiority is also evident in Figure~\ref{fig:cluster_visualization_fault}, where different fault types form distinct clusters. The score exhibits a decreasing trend with an increase in cycles, suggesting that over time, the fault evolves into a higher severity state, leading to the degradation impacting all measurements. Consequently, the clusters begin to overlap, making differentiation more challenging. 

\subsection{Sensor-wise Health Indicator Interpretation}
\begin{figure*}[htbp]
\centerline{\includegraphics[width=0.85\textwidth]{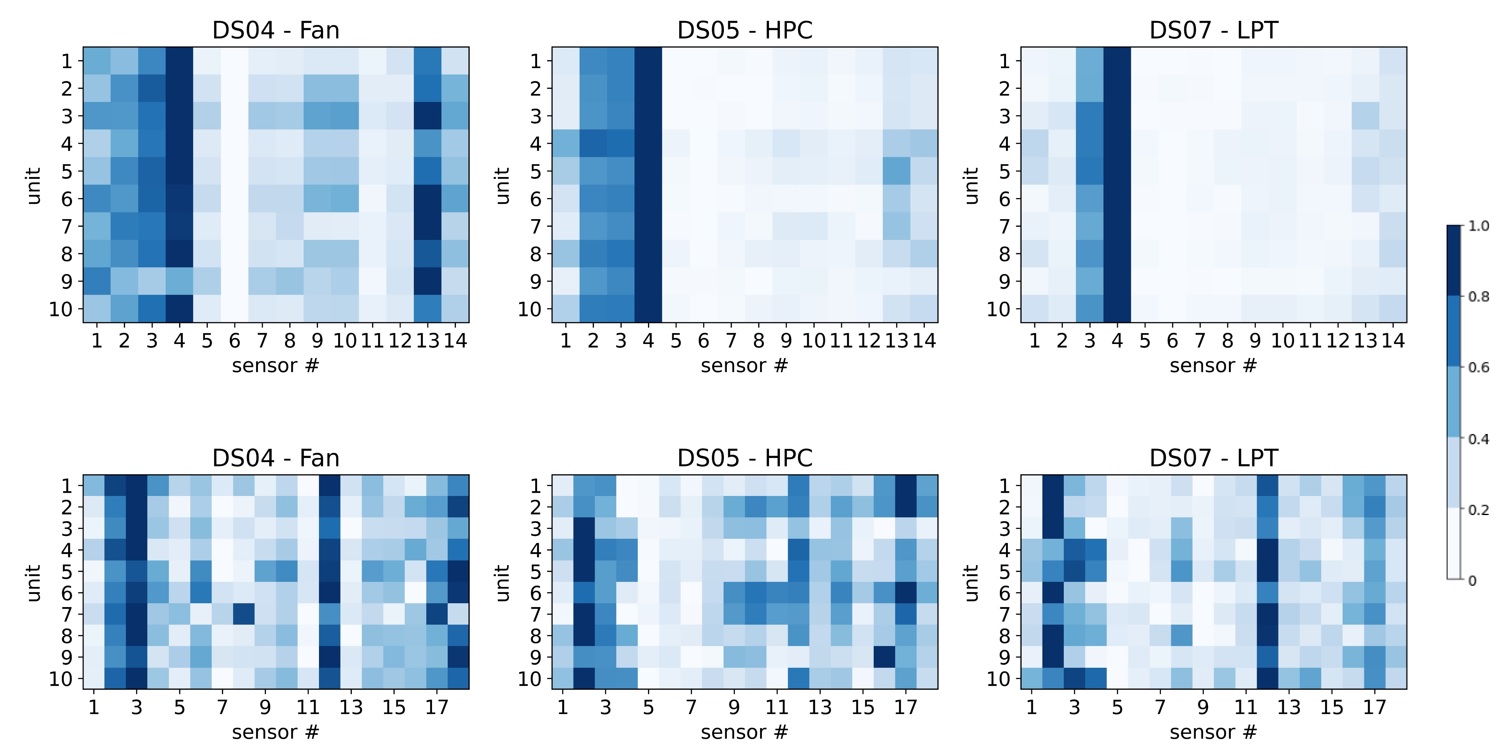}}
\caption{Normalised sensor-wise health indicator $\bm{h}^{\rm S}$ values calculated 10 cycles after the fault detection. The upper figure depicts results using the OC model $\bm{h}^{\rm \rm S\text{-}OC}$, while the lower  figure represents the AE model $\bm{h}^{\rm \rm S\text{-}AE}$.   
In the AE model, sensor 1 to 14 represent the sensor readings $\bm{x}$, while sensors 15 to 18 correspond to operating condition descriptors $\bm{w}$.}
\label{fig:activation_map_oc_ae}
\end{figure*}

\begin{figure*}[htbp]
\centerline{\includegraphics[width=0.85\textwidth]{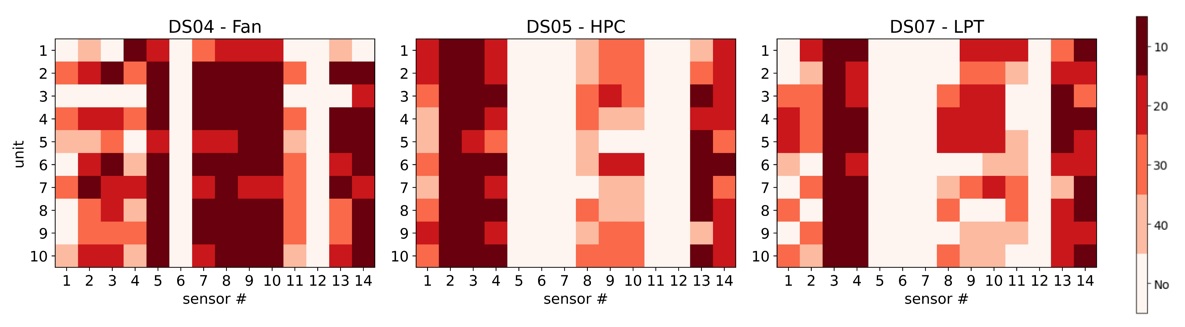}}
\caption{Sensors that are triggered using $\bm{h}^{\rm S-{OC}}$ with $n_c=10, 20, 30, 40$ cycles after the first fault detection. Darker colors indicate earlier sensor triggering. The darkest color signifies triggering within 10 cycles after fault detection. The lightest color, labeled as "No" indicates no triggering up to $n_0+40$ cycles.}
\label{fig:segmentation_over_time}
\end{figure*}

The sensor-wise health indicators $\bm{h}^{\rm S}$ are displayed in Figure~\ref{fig:activation_map_oc_ae} at 10 cycles after fault detection. The residuals are normalised for each unit. In the case of the OC model, a larger number of sensors have high residuals when a fan fault occurs, compared to HPC or LPT faults. Since the fan is the first component in the engine system, most downstream components are affected when a fault arises in this particular component. 

In DS05 with an HPC fault, high residuals are observed in the temperature measurements S2-T30 (temperature at the HPC outlet), S3-T48 (temperature at the HPT outlet), and S4-T50 (temperature at the LPT outlet). Conversely, in DS07 with an LPT fault, high residuals are concentrated only in S3-T48 and S4-T50. It is worth noting that sensor S3-T48, which measures total temperature at the HPT outlet, also serves as the inlet to the LPT.
Together with S4-T50, these two sensors effectively measure the input and output temperature of the faulty component, which is the LPT. To differentiate between the HPC fault in DS05 and the LPT fault in DS07, the key sensor is S2-T30 which measures the total temperature at the HPC outlet, which is the fault component of DS05. The sensor S2-T30 exhibits a high residual only in the case of the HPC fault in DS05 but not in the LPT fault in DS07. 

For the AE model, interpretation becomes more challenging as multiple sensors have high residuals simultaneously, and there is a higher degree of variation between engines. 
This highlights the advantage of the OC model, as it provides more refined residuals that are easier to relate to the physical system. The OC model offers a clearer and more straightforward understanding of the deviations from normal behavior.

In Figure~\ref{fig:segmentation_over_time}, we depict the sensors that have detected faults at different cycle numbers using the sensor-wise health indicator of the OC model $\bm{h}^{\rm S\text{-}OC}$. We have not provided the same figure for the AE model due to excessive variation in sensor activation, which complicates interpretation. This figure is relevant to understanding the evolution of triggered sensors. A triggered sensor is a sensor that has a sensor-wise residual higher than the pre-defined threshold as discussed in Section~\ref{sec:fault_detection}.
Darker colors indicate earlier sensor-wise fault detection. At 10 cycles after the first fault detection, the triggering pattern resembles that shown in Figure~\ref{fig:activation_map_oc_ae}, wherein health indicators with high values are triggered.   

With this figure, it becomes possible to track the evolution of faulty components over time. For instance, in the case of an HPC fault, initially, sensors S2-T30, S3-T48, and S4-T50 are predominantly triggered. As the fault progresses, deviations are observed in S13-$\rm {Nc}$ (physical core speed) and S14-$\rm {Wf}$ (fuel flow), indicating that the fault begins to impact the burner and the shaft. Towards the end-of-life, sensors S8-P24 (pressure at LPC outlet), S9-Ps30 (pressure at HPC outlet), and S10-P40 (pressure at burner outlet) situated around the burner and HPC demonstrate further deterioration. 

This figure can also be used to differentiate between the HPC fault and the LPT fault. While the sensor S2-T30 is triggered in both faults, it exhibits an earlier trigger in the HPC fault compared to the LPT fault because the sensor S2-T30 measures the total temperature at the HPC outlet. In the case of the HPC fault, the impact on the sensor S2-T30 is direct and immediate. However, in the LPT fault, the impact on S2-T30 reading becomes noticeable only after a longer duration, typically 20 or 30 cycles after fault detection.

\section{Conclusion}
\label{sec:conclusion}
In this study, we performed  a comparative analysis of two residual-based models, specifically the AE and OC model, for the purpose of fault detection. We constructed two types of health indicators from residuals: a univariate aggregated health indicator and multivariate sensor-wise health indicators. Our framework was applied to three sub-datasets from  N-CMAPSS, each presenting different fault types in different engine components. 

The results demonstrated  that the sensor-wise health indicator outperformed the aggregated health indicator in terms of fault detection performance. Furthermore, the health indicators obtained using the OC model exhibited superior fault separation capabilities. It effectively highlighted the sensors triggering  and could be directly linked to specific fault components. This research highlights an alternative residual model that surpasses the commonly used AE models in both fault detection and segmentation. It not only demonstrates superior performance but also provides more meaningful health indicators.

As a future direction, it is essential to evaluate the proposed approaches on other systems that exhibit different fault evolution behavior. Additionally, it is important to consider systems with higher variability in terms of operating conditions and their impact on fault evolution. By testing the approaches across a diverse range of systems, we can ensure their effectiveness and applicability in various real-world scenarios. 

Furthermore, the prediction of remaining useful life could be achieved using a two-stage approach, where the prediction begins only after the fault is detected. Another potential avenue involves evaluating the evolution of fault patterns by analyzing factors like the sequence of sensor triggers. Lastly, it would be beneficial to explore different fault detection architectures, such as recurrent neural networks and variational autoencoders.

\section*{Acknowledgment}
This work is part of a project that is financially supported by the Swiss Federal Office of Energy.







\begin{thebibliography}{}

\bibitem [\protect \citeauthoryear {%
Arias~Chao%
, Kulkarni%
, Goebel%
\BCBL {}\ \BBA {} Fink%
}{%
Arias~Chao%
\ \protect \BOthers {.}}{%
{\protect \APACyear {2019}}%
}]{%
chao2019hybrid}
\APACinsertmetastar {%
chao2019hybrid}%
\begin{APACrefauthors}%
Arias~Chao, M.%
, Kulkarni, C.%
, Goebel, K.%
\BCBL {}\ \BBA {} Fink, O.%
\end{APACrefauthors}%
\unskip\
\newblock
\APACrefYearMonthDay{2019}{}{}.
\newblock
{\BBOQ}\APACrefatitle {Hybrid deep fault detection and isolation: Combining
  deep neural networks and system performance models} {Hybrid deep fault
  detection and isolation: Combining deep neural networks and system
  performance models}.{\BBCQ}
\newblock
\APACjournalVolNumPages{International Journal of Prognostics and Health
  Management}{}{}{}.
\PrintBackRefs{\CurrentBib}

\bibitem [\protect \citeauthoryear {%
Arias~Chao%
, Kulkarni%
, Goebel%
\BCBL {}\ \BBA {} Fink%
}{%
Arias~Chao%
\ \protect \BOthers {.}}{%
{\protect \APACyear {2021}}%
}]{%
arias2021aircraft}
\APACinsertmetastar {%
arias2021aircraft}%
\begin{APACrefauthors}%
Arias~Chao, M.%
, Kulkarni, C.%
, Goebel, K.%
\BCBL {}\ \BBA {} Fink, O.%
\end{APACrefauthors}%
\unskip\
\newblock
\APACrefYearMonthDay{2021}{}{}.
\newblock
{\BBOQ}\APACrefatitle {Aircraft engine run-to-failure dataset under real flight
  conditions for prognostics and diagnostics} {Aircraft engine run-to-failure
  dataset under real flight conditions for prognostics and diagnostics}.{\BBCQ}
\newblock
\APACjournalVolNumPages{Data}{6}{1}{5}.
\PrintBackRefs{\CurrentBib}

\bibitem [\protect \citeauthoryear {%
Darrah%
, L{\"o}vberg%
, Frank%
, Biswas%
\BCBL {}\ \BBA {} Quinones-Gruiero%
}{%
Darrah%
\ \protect \BOthers {.}}{%
{\protect \APACyear {2022}}%
}]{%
darrah2022developing}
\APACinsertmetastar {%
darrah2022developing}%
\begin{APACrefauthors}%
Darrah, T.%
, L{\"o}vberg, A.%
, Frank, J.%
, Biswas, G.%
\BCBL {}\ \BBA {} Quinones-Gruiero, M.%
\end{APACrefauthors}%
\unskip\
\newblock
\APACrefYearMonthDay{2022}{}{}.
\newblock
{\BBOQ}\APACrefatitle {Developing Deep Learning Models for System Remaining
  Useful Life Predictions: Application to Aircraft Engines} {Developing deep
  learning models for system remaining useful life predictions: Application to
  aircraft engines}.{\BBCQ}
\newblock
\BIn{} \APACrefbtitle {Annual Conference of the PHM Society} {Annual conference
  of the phm society}\ (\BVOL~14).
\PrintBackRefs{\CurrentBib}

\bibitem [\protect \citeauthoryear {%
Fink%
\ \protect \BOthers {.}}{%
Fink%
\ \protect \BOthers {.}}{%
{\protect \APACyear {2020}}%
}]{%
fink2020potential}
\APACinsertmetastar {%
fink2020potential}%
\begin{APACrefauthors}%
Fink, O.%
, Wang, Q.%
, Svensen, M.%
, Dersin, P.%
, Lee, W\BHBI J.%
\BCBL {}\ \BBA {} Ducoffe, M.%
\end{APACrefauthors}%
\unskip\
\newblock
\APACrefYearMonthDay{2020}{}{}.
\newblock
{\BBOQ}\APACrefatitle {Potential, challenges and future directions for deep
  learning in prognostics and health management applications} {Potential,
  challenges and future directions for deep learning in prognostics and health
  management applications}.{\BBCQ}
\newblock
\APACjournalVolNumPages{Engineering Applications of Artificial
  Intelligence}{92}{}{103678}.
\PrintBackRefs{\CurrentBib}

\bibitem [\protect \citeauthoryear {%
Frederick%
, DeCastro%
\BCBL {}\ \BBA {} Litt%
}{%
Frederick%
\ \protect \BOthers {.}}{%
{\protect \APACyear {2007}}%
}]{%
frederick2007user}
\APACinsertmetastar {%
frederick2007user}%
\begin{APACrefauthors}%
Frederick, D\BPBI K.%
, DeCastro, J\BPBI A.%
\BCBL {}\ \BBA {} Litt, J\BPBI S.%
\end{APACrefauthors}%
\unskip\
\newblock
\APACrefYearMonthDay{2007}{}{}.
\newblock
\APACrefbtitle {User's guide for the commercial modular aero-propulsion system
  simulation (C-MAPSS)} {User's guide for the commercial modular
  aero-propulsion system simulation (c-mapss)}\ \APACbVolEdTR{}{\BTR{}}.
\PrintBackRefs{\CurrentBib}

\bibitem [\protect \citeauthoryear {%
Guo%
, Yu%
, Duan%
, Gao%
\BCBL {}\ \BBA {} Zhang%
}{%
Guo%
\ \protect \BOthers {.}}{%
{\protect \APACyear {2022}}%
}]{%
guo2022unsupervised}
\APACinsertmetastar {%
guo2022unsupervised}%
\begin{APACrefauthors}%
Guo, L.%
, Yu, Y.%
, Duan, A.%
, Gao, H.%
\BCBL {}\ \BBA {} Zhang, J.%
\end{APACrefauthors}%
\unskip\
\newblock
\APACrefYearMonthDay{2022}{}{}.
\newblock
{\BBOQ}\APACrefatitle {An unsupervised feature learning based health indicator
  construction method for performance assessment of machines} {An unsupervised
  feature learning based health indicator construction method for performance
  assessment of machines}.{\BBCQ}
\newblock
\APACjournalVolNumPages{Mechanical Systems and Signal
  Processing}{167}{}{108573}.
\PrintBackRefs{\CurrentBib}

\bibitem [\protect \citeauthoryear {%
Kingma%
\ \BBA {} Ba%
}{%
Kingma%
\ \BBA {} Ba%
}{%
{\protect \APACyear {2014}}%
}]{%
kingma2014adam}
\APACinsertmetastar {%
kingma2014adam}%
\begin{APACrefauthors}%
Kingma, D\BPBI P.%
\BCBT {}\ \BBA {} Ba, J.%
\end{APACrefauthors}%
\unskip\
\newblock
\APACrefYearMonthDay{2014}{}{}.
\newblock
{\BBOQ}\APACrefatitle {Adam: A method for stochastic optimization} {Adam: A
  method for stochastic optimization}.{\BBCQ}
\newblock
\APACjournalVolNumPages{arXiv preprint arXiv:1412.6980}{}{}{}.
\PrintBackRefs{\CurrentBib}

\bibitem [\protect \citeauthoryear {%
Lei%
\ \protect \BOthers {.}}{%
Lei%
\ \protect \BOthers {.}}{%
{\protect \APACyear {2018}}%
}]{%
lei2018machinery}
\APACinsertmetastar {%
lei2018machinery}%
\begin{APACrefauthors}%
Lei, Y.%
, Li, N.%
, Guo, L.%
, Li, N.%
, Yan, T.%
\BCBL {}\ \BBA {} Lin, J.%
\end{APACrefauthors}%
\unskip\
\newblock
\APACrefYearMonthDay{2018}{}{}.
\newblock
{\BBOQ}\APACrefatitle {Machinery health prognostics: A systematic review from
  data acquisition to RUL prediction} {Machinery health prognostics: A
  systematic review from data acquisition to rul prediction}.{\BBCQ}
\newblock
\APACjournalVolNumPages{Mechanical systems and signal
  processing}{104}{}{799--834}.
\PrintBackRefs{\CurrentBib}

\bibitem [\protect \citeauthoryear {%
L{\"o}vberg%
}{%
L{\"o}vberg%
}{%
{\protect \APACyear {2021}}%
}]{%
lovberg2021remaining}
\APACinsertmetastar {%
lovberg2021remaining}%
\begin{APACrefauthors}%
L{\"o}vberg, A.%
\end{APACrefauthors}%
\unskip\
\newblock
\APACrefYearMonthDay{2021}{}{}.
\newblock
{\BBOQ}\APACrefatitle {Remaining Useful Life Prediction of Aircraft Engines
  with Variable Length Input Sequences} {Remaining useful life prediction of
  aircraft engines with variable length input sequences}.{\BBCQ}
\newblock
\BIn{} \APACrefbtitle {Proceedings of the Annual Conference of the PHM
  Society.} {Proceedings of the annual conference of the phm society.}
\PrintBackRefs{\CurrentBib}

\bibitem [\protect \citeauthoryear {%
Michau%
, Hu%
, Palm{\'e}%
\BCBL {}\ \BBA {} Fink%
}{%
Michau%
\ \protect \BOthers {.}}{%
{\protect \APACyear {2020}}%
}]{%
michau2020feature}
\APACinsertmetastar {%
michau2020feature}%
\begin{APACrefauthors}%
Michau, G.%
, Hu, Y.%
, Palm{\'e}, T.%
\BCBL {}\ \BBA {} Fink, O.%
\end{APACrefauthors}%
\unskip\
\newblock
\APACrefYearMonthDay{2020}{}{}.
\newblock
{\BBOQ}\APACrefatitle {Feature learning for fault detection in high-dimensional
  condition monitoring signals} {Feature learning for fault detection in
  high-dimensional condition monitoring signals}.{\BBCQ}
\newblock
\APACjournalVolNumPages{Proceedings of the Institution of Mechanical Engineers,
  Part O: Journal of Risk and Reliability}{234}{1}{104--115}.
\PrintBackRefs{\CurrentBib}

\bibitem [\protect \citeauthoryear {%
Michau%
, Palm{\'e}%
\BCBL {}\ \BBA {} Fink%
}{%
Michau%
\ \protect \BOthers {.}}{%
{\protect \APACyear {2017}}%
}]{%
michau2017deep}
\APACinsertmetastar {%
michau2017deep}%
\begin{APACrefauthors}%
Michau, G.%
, Palm{\'e}, T.%
\BCBL {}\ \BBA {} Fink, O.%
\end{APACrefauthors}%
\unskip\
\newblock
\APACrefYearMonthDay{2017}{}{}.
\newblock
{\BBOQ}\APACrefatitle {Deep feature learning network for fault detection and
  isolation} {Deep feature learning network for fault detection and
  isolation}.{\BBCQ}
\newblock
\BIn{} \APACrefbtitle {Annual Conference of the PHM Society} {Annual conference
  of the phm society}\ (\BVOL~9).
\PrintBackRefs{\CurrentBib}

\bibitem [\protect \citeauthoryear {%
Pan%
, Meng%
, Chen%
, Gao%
\BCBL {}\ \BBA {} Shi%
}{%
Pan%
\ \protect \BOthers {.}}{%
{\protect \APACyear {2020}}%
}]{%
pan2020two}
\APACinsertmetastar {%
pan2020two}%
\begin{APACrefauthors}%
Pan, Z.%
, Meng, Z.%
, Chen, Z.%
, Gao, W.%
\BCBL {}\ \BBA {} Shi, Y.%
\end{APACrefauthors}%
\unskip\
\newblock
\APACrefYearMonthDay{2020}{}{}.
\newblock
{\BBOQ}\APACrefatitle {A two-stage method based on extreme learning machine for
  predicting the remaining useful life of rolling-element bearings} {A
  two-stage method based on extreme learning machine for predicting the
  remaining useful life of rolling-element bearings}.{\BBCQ}
\newblock
\APACjournalVolNumPages{Mechanical Systems and Signal
  Processing}{144}{}{106899}.
\PrintBackRefs{\CurrentBib}

\bibitem [\protect \citeauthoryear {%
Rausch%
, Goebel%
, Eklund%
\BCBL {}\ \BBA {} Brunell%
}{%
Rausch%
\ \protect \BOthers {.}}{%
{\protect \APACyear {2007}}%
}]{%
rausch2007integrated}
\APACinsertmetastar {%
rausch2007integrated}%
\begin{APACrefauthors}%
Rausch, R\BPBI T.%
, Goebel, K\BPBI F.%
, Eklund, N\BPBI H.%
\BCBL {}\ \BBA {} Brunell, B\BPBI J.%
\end{APACrefauthors}%
\unskip\
\newblock
\APACrefYearMonthDay{2007}{01}{}.
\newblock
{\BBOQ}\APACrefatitle {{Integrated in-Flight Fault Detection and Accommodation:
  A Model-Based Study}} {{Integrated in-Flight Fault Detection and
  Accommodation: A Model-Based Study}}.{\BBCQ}
\newblock
\APACjournalVolNumPages{Journal of Engineering for Gas Turbines and
  Power}{129}{4}{962-969}.
\newblock
\begin{APACrefURL} \url{https://doi.org/10.1115/1.2720517} \end{APACrefURL}
\newblock
\begin{APACrefDOI} \doi{10.1115/1.2720517} \end{APACrefDOI}
\PrintBackRefs{\CurrentBib}

\bibitem [\protect \citeauthoryear {%
Reddy%
, Sarkar%
, Venugopalan%
\BCBL {}\ \BBA {} Giering%
}{%
Reddy%
\ \protect \BOthers {.}}{%
{\protect \APACyear {2016}}%
}]{%
reddy2016anomaly}
\APACinsertmetastar {%
reddy2016anomaly}%
\begin{APACrefauthors}%
Reddy, K\BPBI K.%
, Sarkar, S.%
, Venugopalan, V.%
\BCBL {}\ \BBA {} Giering, M.%
\end{APACrefauthors}%
\unskip\
\newblock
\APACrefYearMonthDay{2016}{}{}.
\newblock
{\BBOQ}\APACrefatitle {Anomaly detection and fault disambiguation in large
  flight data: A multi-modal deep auto-encoder approach} {Anomaly detection and
  fault disambiguation in large flight data: A multi-modal deep auto-encoder
  approach}.{\BBCQ}
\newblock
\BIn{} \APACrefbtitle {Annual Conference of the PHM Society} {Annual conference
  of the phm society}\ (\BVOL~8).
\PrintBackRefs{\CurrentBib}

\bibitem [\protect \citeauthoryear {%
Rousseeuw%
}{%
Rousseeuw%
}{%
{\protect \APACyear {1987}}%
}]{%
rousseeuw1987silhouettes}
\APACinsertmetastar {%
rousseeuw1987silhouettes}%
\begin{APACrefauthors}%
Rousseeuw, P\BPBI J.%
\end{APACrefauthors}%
\unskip\
\newblock
\APACrefYearMonthDay{1987}{}{}.
\newblock
{\BBOQ}\APACrefatitle {Silhouettes: a graphical aid to the interpretation and
  validation of cluster analysis} {Silhouettes: a graphical aid to the
  interpretation and validation of cluster analysis}.{\BBCQ}
\newblock
\APACjournalVolNumPages{Journal of computational and applied
  mathematics}{20}{}{53--65}.
\PrintBackRefs{\CurrentBib}

\bibitem [\protect \citeauthoryear {%
Saufi%
, Ahmad%
, Leong%
\BCBL {}\ \BBA {} Lim%
}{%
Saufi%
\ \protect \BOthers {.}}{%
{\protect \APACyear {2019}}%
}]{%
saufi2019challenges}
\APACinsertmetastar {%
saufi2019challenges}%
\begin{APACrefauthors}%
Saufi, S\BPBI R.%
, Ahmad, Z\BPBI A\BPBI B.%
, Leong, M\BPBI S.%
\BCBL {}\ \BBA {} Lim, M\BPBI H.%
\end{APACrefauthors}%
\unskip\
\newblock
\APACrefYearMonthDay{2019}{}{}.
\newblock
{\BBOQ}\APACrefatitle {Challenges and opportunities of deep learning models for
  machinery fault detection and diagnosis: A review} {Challenges and
  opportunities of deep learning models for machinery fault detection and
  diagnosis: A review}.{\BBCQ}
\newblock
\APACjournalVolNumPages{Ieee Access}{7}{}{122644--122662}.
\PrintBackRefs{\CurrentBib}

\end{thebibliography}
\end{document}